# Magnetic resonance and microwave resistance modulation in van der Waals Ferrimagnet $Mn_3Si_2Te_6$

Miuko Tanaka[1]*, Abdul Ahad[1], Darius-Alexandru Deaconu[2], Varun Shah[2], Daisuke Nishio-Hamane[1], Tomohiro Ishii[1], Masayuki Hashisaka[1], Shoya Sakamoto[1], Shinji Miwa[1], Mohammad Saeed Bahramy[2], Toshiya Ideue*[1]

[1]Institute for Solid State Physics, The University of Tokyo, Kashiwa-shi, Japan
[2]Department of Physics and Astronomy, The University of Manchester, Oxford Road, Manchester M13 9PL, United Kingdom

**ABSTRACT**

Unconventional magnetoresistance is a fascinating quantum phenomenon that continues to draw significant interest in condensed matter physics. $Mn_3Si_2Te_6$ has emerged as such a novel magnetoresistance material, notable for its largest MR exceeding conventional colossal magnetoresistance materials and pronounced directional anisotropy. Despite extensive research, the mechanisms driving MR in $Mn_3Si_2Te_6$ remain elusive [1-4]. In this work, we explore the magnetic resonance of $Mn_3Si_2Te_6$ and observe a reduced g-factor for magnetic fields applied along the crystalline c-axis compared to the ab-plane, indicating a substantial orbital magnetization contribution along the c-axis. Furthermore, we detect resistance modulation under resonance conditions, suggesting that MR in $Mn_3Si_2Te_6$ is sensitive to the out-of-the plane spin polarization. These findings shed new light on the role of orbital magnetic moment in $Mn_3Si_2Te_6$, offering a deeper understanding of the interplay between spin, orbital and lattice degrees of freedom of electrons.

## I. INTRODUCTION.

Layered intercalated magnets are known to exhibit versatile exotic magnetic structures and resultant emergent magnetic properties [5-8], attracting great attention in recent years. Recently, the layered ferrimagnetic material $Mn_3Si_2Te_6$ has been found to exhibit the largest MR among known materials, as well as unconventional anisotropy, which is distinct from known CMR materials [1-4, 9-35]. It is an easy-plane (thought to be isotropic within ab-plane) ferrimagnet, which shows MR only when the magnetic field is applied along the hard c-axis (fig.1a) [1-4]. The band structure and manganese mono-valency ($Mn^{2+}$) rule out conventional CMR mechanisms. Although various theoretical models have been proposed to explain its MR, a definitive understanding remains elusive.

The first proposed mechanism was a spin-rotation-induced insulator-to-metal transition, in which the authors considered a fully aligned ferrimagnetic spin configuration along the c-axis and attributed MR to the removal of Mn–Te orbital hybridization-driven band degeneracy (i.e., L ∥ S) [1]. More recently, an alternative explanation involving chiral orbital currents (COC) has been proposed as the driving mechanism [3]. However, no direct experimental evidence for COC has been reported to date. Additionally, alternative explanations such as heating effects under large DC currents and the influence of magnetic domain structures in bulk samples have also been suggested as factors complicating intrinsic MR observations [2].

$Mn_3Si_2Te_6$ is a 2D van der Waals material that is inherently anisotropic. However, it exhibits counterintuitive behavior by being more conductive in the out-of-plane than in plane direction, distinguishing it from most other 2D magnetic materials. Furthermore, neutron diffraction and DC magnetization measurements revealed the non-saturation of magnetization even up to applied field of 13 T while the MR could be observed at ~ 3T (at 10 K), posing a fundamental puzzle [1].

The available literature suggests that the interplay between spin canting and orbital moments from the Te sublattice plays a crucial role. Spin dynamics in $Mn_3Si_2Te_6$ have not yet been extensively studied.

*Contact author: miukot@issp.u-tokyo.ac.jp

Since magnetic resonance measurement can prove magnetic anisotropy, spin dissipation processes, and g-factor [37-39], it has been widely applied to study conventional CMR materials [28-35]. It can therefore offer valuable insights into the microscopic mechanism of the orbital-driven unconventional MR in $Mn_3Si_2Te_6$. In this study, we measure the temperature-dependent magnetic resonance signal. Our analysis provides a quantitative estimate of the canting angle and the renormalized Lande g-factor, evidencing the presence of orbital moments along the c-axis. Furthermore, we found that the unique unconventional MR enables a way to modulate the transport properties under the magnetic resonance.

## II. EXPERIMENTAL METHOD

The crystals used in this study were commercially outsourced (2D Semiconductors, USA). The chemical composition of the sample is analyzed using scanning electron microscopy-energy dispersive X-ray spectroscopy (SEM-EDX; IT100, JEOL) at 15 kV and 0.8 nA.

For electrical transport measurements, we used the Quantum Design Physical Property Measurement (QD PPMS) system to record the temperature and field dependent resistivity ($\rho_{ab}$) using four probe contact methods. Keithley 2612A source meter and Agilent 34410 multimeter were employed to source the current and record the voltages, respectively.

All the magnetization measurements were carried out by utilizing the 7 T Quantum Design SQUID VSM. DC magnetization vs temperature (5 to 300 K) measurements at various applied magnetic fields were recorded under field cooled (FC) protocol. The isothermal M-H hysteresis loop was measured at 10 K in both configurations i.e., $H$ || ab and $H$ || c.

For the magnetic resonance measurement, well characterized crystals with sizes ~ 4 x 3 mm were placed on the coplanar wave guide (CPW) patterned on the PCB sample holder with characteristic impedance of 50 Ω. 9 T QD PPMS was used to control the magnetic field and the temperatures during measurements. Vector network analyzer was used to source and measure $S_{21}$ parameter.

The bulk electronic structure calculations for the hexagonal structure of $Mn_3Si_2Te_6$ are performed within the density functional theory (DFT) using the Perdew-Burke-Ernzerhof exchange-correlation functional [40] as implemented in Vienna Ab initio Simulation Package (VASP) [41]. Relativistic effects, including spin-orbit coupling, were fully taken into account. The corresponding Brillouin zone was sampled by 8 ×8 ×4 Γ-centered k-mesh. Spin and orbital angular momenta were computed using accurate tight-binding Hamiltonians downfolded from the DFT calculations through Wannier interpolation with the Mn-d orbitals and Te-p orbitals as projection centers [42].

## III. RESULTS and DISCUSSION

### A. Sample characterization

The crystals are first characterized by Energy Dispersive X-ray Spectroscopy (EDX). Results indicated the atomic concentration ratio as Mn/Si : 1.84 and Mn/Te : 0.467, confirming the composition close to $Mn_3Si_2Te_6$ (Fig.1b, [43]).

In the resistivity, a well resolved characteristic hump at $T$c ~ 78 K is observed, which corresponds to its magnetic ordering temperature (Fig.1c). Moreover, the MR behavior is also observed (Fig.1d). All the observations are consistent with the previous reports [1-3].

Next, to check the static magnetic properties, magnetization measurements are carried out as a function of temperature at various applied magnetic fields under field cooled (FC) protocol (Fig.1e, f). Magnetization starts to rise at $T_C$=78 K, confirming the critical temperature. The isothermal M-H hysteresis loop was measured at 10 K in both configurations i.e., $H$||ab and $H$||c. For $H$||ab, the magnetization saturates at around 60 mT, while for $H$||c the magnetization does not saturate up to 3.5 T, which indicates the presence of easy-plane anisotropy, consistent with previous reports [1,36].

### B. Magnetic resonance

After confirming properties consistent with previous research, we study the magnetic resonance in this material. A crystal with the size of ~4×3mm is placed on the coplanar wave guide (CPW) patterned on FR4 PCB substrate with characteristic impedance of 50 Ω. The wave guide PCB is connected to a 50 Ω coaxial cable and placed inside a PPMS to control the temperature and magnetic field. The $S_{21}$ parameter is measured using a vector network analyzer. The microwave power at the CPW under the sample is approximately -34dBm, generating a microwave magnetic field of about $10^{-6}$ T applied to the sample.

Figure 2 shows the magnetic resonance mode in three different magnetic field configurations at 70 K. When the magnetic field is applied in-plane (Fig2a, b), the magnon mode exhibits a monotonic behavior, which can be well described by the Kittel formula [44], Eq. (1) [45]:

$$\frac{\omega^2}{\gamma^2} = \mu_0 H\left(\mu_0 H + B_K + \mu_0 M\right). \qquad Eq\ (1)$$

Here, $B_K$ is anisotropy field, $M$ is saturation magnetization, and $\gamma = g\mu_B$ is the gyromagnetic ratio, g is the g-factor, and $\mu_B$ is the Bohr magneton. The resonance mode is well fitted by the parameters, $B_K$ =4 T, $\mu_0 M$=0.18 T, and γ =28 GHz/T. The extracted saturation magnetization corresponds well with the magnetization measurement, confirming $M$=1.6$\mu_B$/Mn [1,36]. The anisotropy field has the same order of magnitude as the values obtained from magnetization measurements, which increases and saturates in the range of 5.7–8 T at low temperatures (Fig.3) [1].

Fig.2f presents the resonance mode under $H$||c-axis. Unlike the $H$||ab-plane case, the resonance mode is non-monotonic. This behavior is known to appear when the applied magnetic field is not parallel to the magnetization [44,46-47]. The resonance mode can be described in the Kittel formalism, as follows, Eq (2) [45]:

$$\left(\frac{\omega}{\gamma}\right)^2 = [\mu_0 H\cos\theta_0 + B_K\cos^2\theta_0 - \mu_0 M\cos\theta_0] \times [\mu_0 H\cos\theta_0 + B_K\cos 2\theta_0 - \mu_0 M\cos 2\theta_0] + \mu_0 H_x(\mu_0 H_x + B_K\sin\theta_0), \qquad Eq. (2)$$

where $\theta_0$ is the angle between the magnetization direction and the c-axis and is determined by $\cos\theta_0 = \frac{\mu_0 H}{\mu_0 M - B_K}$.

The resonance mode fits well with this model, yielding comparable values for $B_K$ and $\mu_0 M$ as obtained for $H$||ab-plane. However, a notable difference is observed: the resonance mode for $H$||c-axis cannot be fitted by γ =28 GHz/T but require smaller γ~18 GHz/T. Since gyromagnetic ratio is proportional to the g-factor, and γ =28GHz/T corresponds to g-factor of 2 for bare electron, this result suggests that the effective g-factor is reduced for $H$||c-axis.

### C. Temperature dependence and orbital magnetic moment

Figures 3a and 3b show the temperature dependence of the fitting parameters. For in-plane $H$ configurations, the g-factor remains constant at g=2 across all temperatures. However, for $H$||c, the g-factor is reduced to approximately 1.3~1.4 and remains independent of temperature (Fig.3a). In contrast, the anisotropy field is comparable across all three field configurations (Fig. 3b).

This result suggests the presence of additional magnetic moments along the c-axis. In the presence of spin-orbit coupling, the electron's orbital motion strongly influences its coupling to an external magnetic field. To account for this, the g-factor in a bulk system can be expressed as the sum of two contributions:

$$g = g_s + \frac{1}{\hbar}\left(\langle\uparrow|\hat{L}|\uparrow\rangle - \langle\downarrow|\hat{L}|\downarrow\rangle\right),$$

where $g_s$≈2.0 is the Lande g-factor of a free electron, and L denotes the orbital angular momentum (OAM) parallel to the external magnetic field [48,49].

The electronic band structure, along with the spin and orbital angular momentum, calculated from first principles, are shown for two extreme magnetic configurations in Figure 3 (c-d), (f-g). The former, with the local magnetic moments on Mn oriented along the a-axis, suggest that when H is in-plane, i.e., $|S|=\langle S_x \rangle$, the OAM is quenched ($\langle L_x \rangle=0$), leading to the absence of a significant correction from the spin-orbit interaction energy because <L·S> is approximately equal to zero. The spin-polarized bands near the Fermi level are shown in Fig. 3c, while Fig.3d illustrates the vanishing OAM in this case.

Consequently, the calculations predict no deviation from the free-electron g-factor, consistent with the experimental observation of $g \approx g_s \approx 2.0$.

In contrast, under large $H$||c the local magnetic moments tilt towards the c-axis. In our calculation, we consider the case where all moments are aligned along the c-axis, as the field mainly reorients the local moments without changing their magnitude or quantizing the electronic bands into Landau levels. In this scenario, the spin-polarized bands near the Fermi level exhibit a significant OAM with a net polarization opposite that of the spin majority, quantified as $\langle \uparrow | \boldsymbol{L}_z | \uparrow \rangle \approx -0.6$ (see Fig.3f and g). This leads to a substantial enhancement of spin-orbit coupling energy within the Te-5p bands, thereby significantly modifying the g-factor. More precisely, the orbital contribution is expected to reduce the effective g-factor to: g ≈ 2.0 − 0.6 = 1.4. This value aligns well with the experimental data. In the out-of-plane magnetic field range used in our experiments, the spins are not fully aligned along the c-axis but are canted by approximately 10~20° away from it at $H_{||c}$=6~9 T (see SM). Additional calculations for a finite canting angle of 15° show that the value of $\langle \uparrow | \boldsymbol{L}_z | \uparrow \rangle$ remains essentially unchanged compared with the fully aligned case, further supporting the quantitative consistency between our calculations and the experimental observations (see SM).

Interestingly, this result is also consistent with the Lande g-factor for an electron in a p orbital:

$$g = 1 + \frac{j(j+1) + s(s+1) - l(l+1)}{2j(j+1)} = 1.33$$

where s = 1/2, l = 1, and j = l + s = 3/2 are the spin, orbital, and total angular momentum quantum numbers, respectively. This further underscores the critical role of Te-5p orbitals in the magnetic response of $Mn_3Si_2Te_6$ to an external field and their major contribution to the effective g-factor of the system. In our calculations, the application of an out-of-plane magnetic field induces a finite orbital magnetic moment, which leads to an upward shift of the valence band energy, causing it to cross the Fermi level. This leads to a metal to insulator transition, and accounts for the unconventional magnetoresistance observed in this material.

### D. Resistance modulation under magnetic resonance

Since $Mn_3Si_2Te_6$ shows the unique unconventional MR sensitive to the c-axis spin component, resistance modulation is expected when spin resonance is excited. Microwave-induced resistance modulation has been conventionally studied in magnetic metals and semiconductors. For example, in ferromagnetic metals, it is known that magnetization precession under microwave excitation can modulate the resistance via anisotropic magnetoresistance effect [49-51]. In organic semiconductors, electrically detected magnetic resonance (EDMR) is reported, which originates from the difference of the recombination rate in spin singlet and triplet pairs formed between the local spins and conduction electrons [52-54]. Microwave-induced resistance modulation in $Mn_3Si_2Te_6$ as well as conventional CMR magnets has not yet been reported so far and is of great interest from the viewpoint of its origin and the possibility of new functionalities.

To investigate this, we measure the resistance change, which is proportional to the microwave (MW) amplitude, using lock-in detection (Fig.4a). The sample is placed on the CPW, with two-terminal electrical contacts made across the CPW. A constant DC voltage is applied to the sample while the current is monitored as both microwave and DC magnetic fields are applied to the CPW. The microwave amplitude is modulated at 40.774 Hz using a voltage-tunable attenuator, and the resulting current modulation at these frequencies is detected with a lock-in amplifier.

Figure 4c shows the current modulation signal as a function of microwave frequency for different out-of-plane DC magnetic fields. A modulation signal is observed near the resonance frequency determined from previous microwave absorption measurements [55]. When the temperature is varied while keeping the magnetic field fixed at 0.4 T (Fig. 4d), the temperature dependence of the modulation signal is consistent with the microwave absorption measurements.

The magnitude of the current modulation is of the order of $2\times10^{-11}$ A, corresponding to a relative resistance change $\Delta R/R$ in between $10^{-5}$ and $10^{-6}$. On the CPW, the applied microwave magnetic field is estimated to be $10^{-6}$ T. This microwave magnetic field tilts the spins and reduces the effective magnetization. Assuming that the microwave magnetic field has a similar effect on resistance as a DC magnetic field of the same magnitude, the expected resistance change due to resonance is of the order of $\Delta R/R \sim 10^{-5}$, which is comparable to the observed value. This suggests that the resistance modulation is closely related to the unconventional MR effect. Note that although the spin tilt due to spin resonance is small, such observable resistance modulation can be realized because of the unique MR behavior.

However, one notable unexpected feature is observed: at low frequencies, the modulation signal exhibits a single peak, while around 11 GHz, it transitions into a peak-dip structure, and at higher frequencies, it forms a single dip. The origin of this crossover in the modulation signal remains unknown and should be further pursued in the future, but it appears to be related to a characteristic timescale of the system, as it consistently occurs around 11 GHz, independent of the magnetic field and temperature (Fig. 4d). Possible timescales responsible for this phenomenon may include carrier scattering time, the dynamics of orbital magnetization, domain motion, or other intrinsic processes.

## IV. CONCLUSIONS

In summary, we investigate the magnetic resonance in $Mn_3Si_2Te_6$, revealing key insights into its spin dynamics and orbital magnetization. Resonance measurements under both in-plane and out-of-plane magnetic fields show that the g-factor is significantly reduced in the presence of an out-of-plane field; this indicates the presence of orbital magnetization along the c-axis. This finding is further supported by theoretical calculations, which demonstrate that the spin-orbit coupling in $Mn_3Si_2Te_6$ strongly influences the Te-5p orbitals, leading to the observed g-factor modification.

In addition, we observed resistance modulation under the magnetic resonance which possibly reflects the unique orbital-driven unconventional MR behavior of this material. The observed modulation of resistance quantitatively matches the expected magnetoresistance behavior, confirming that spin resonance directly affects electrical transport properties. This resistive detection method offers an alternative approach to studying spin dynamics in ferrimagnetic materials, especially in nano flakes, with potential applications in GHz-frequency spintronic devices.

Our results provide further experimental evidence for out-of-plane orbital magnetism in $Mn_3Si_2Te_6$. Given its strong magnetic anisotropy, tunable spin-orbit interactions, and orbital-driven unconventional magnetoresistance, $Mn_3Si_2Te_6$ emerges as a promising candidate for spintronic and magnetoelectric applications, particularly in high-frequency microwave technologies. Future studies in nano flakes or the 2D limit are expected to further explore its potential.

## ACKNOWLEDGMENTS

We acknowledge helpful discussions with Daisuke Nakamura and Mizuki Urai. M.T. acknowledges support from the JSPS KAKENHI (grant no. 23K19026), Murata Science and Education Foundation, and JST, PRESTO (grant no. JPMJPR24H7). M.S.B. acknowledges support from Leverhulme Trust (Grant No. RPG-2023-253). D.A.D. and V.S. had support from the Engineering and Physical Sciences Research Council Standard Research Studentship (DTP) EP/T517823/1 and EP/W5243471/1, respectively.

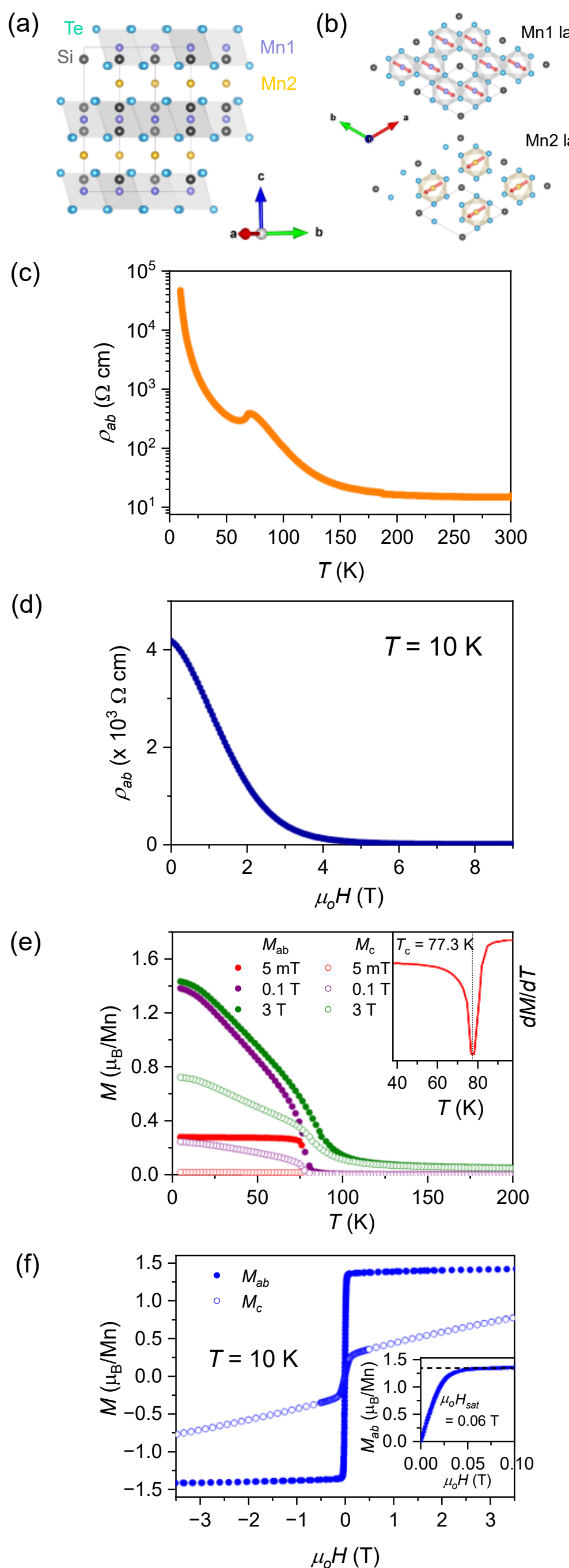

FIG 1. Resistance and magnetization of $Mn_3Si_2Te_6$. (a, b) Crystal structure of $Mn_3Si_2Te_6$ shown from the side (a) and top (b). (c) Four-terminal resistivity as a function of temperature at $H$=0. The current is applied within the ab-plane. (d) Four-terminal resistivity as a function of magnetic field along the c-axis at $T$=10K. The current is applied within the ab-plane. (e) Magnetization as a function of temperature at $\mu_0 H$=5 mT, 0.1 T, and 3 T. Filled circles represent data for $H$||ab-plane, and open circles for $H$||c-axis. The inset shows the temperature derivative of the magnetization at $\mu_0 H$ =5 mT for H||ab-plane. (f) Magnetization as a function of magnetic field applied along the ab-plane (filled circles) and c-axis (open circles) at $T$=10 K. The inset shows a magnified view of the $H$||ab-plane data at low fields.

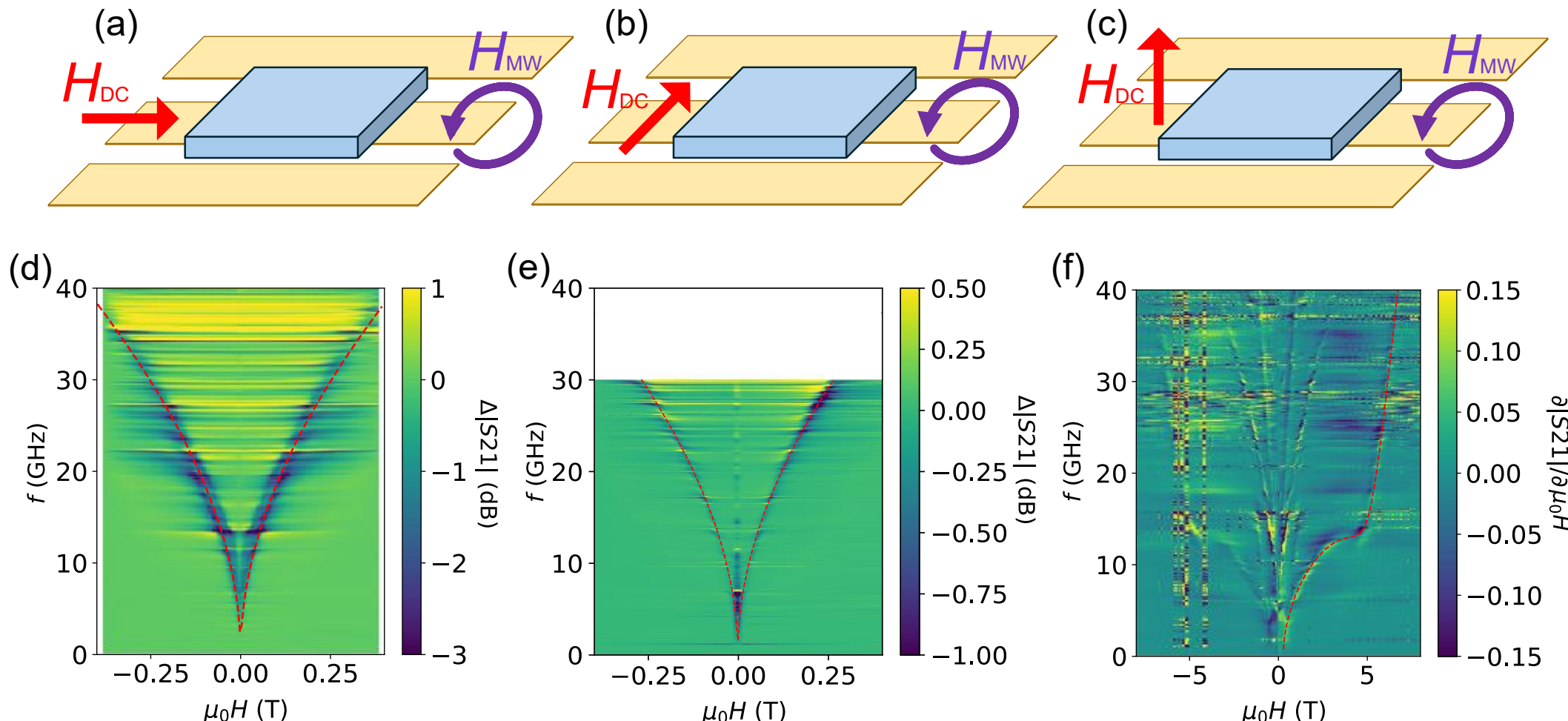

FIG 2. Magnetic resonance in three configurations of static magnetic field. (a) Schematic of the coplanar wave guide (CPW), the sample, and the magnetic field configuration of $H$||ab-plane, $H \perp H_{\mathrm{MW}}$. (b) Magnetic field configuration of $H$||ab-plane, $H$||$H_{\mathrm{MW}}$. (c) Magnetic field configuration of $H$||c-axis. (d) Microwave transmission coefficient |S21| as a function of magnetic field and microwave frequency at 70 K for the setup shown in (a), namely $H$||ab-plane and $H \perp H_{\mathrm{MW}}$. (e) |S21| as a function of magnetic field and microwave frequency for the setup shown in (b), where H||ab-plane and H||HMW. Data measured at 90 K (above the magnetic ordering temperature), where no resonance feature is present, is subtracted from both plots shown in (d) and (e). The dashed red lines represent fits using Eq.1. (f) Derivative with respect to the magnetic field d|S21|/d$\mu_0 H$ as a function of magnetic field and microwave frequency for the setup shown in (c), $H$||c-axis. The dashed red lines represent fits using Eq.2.

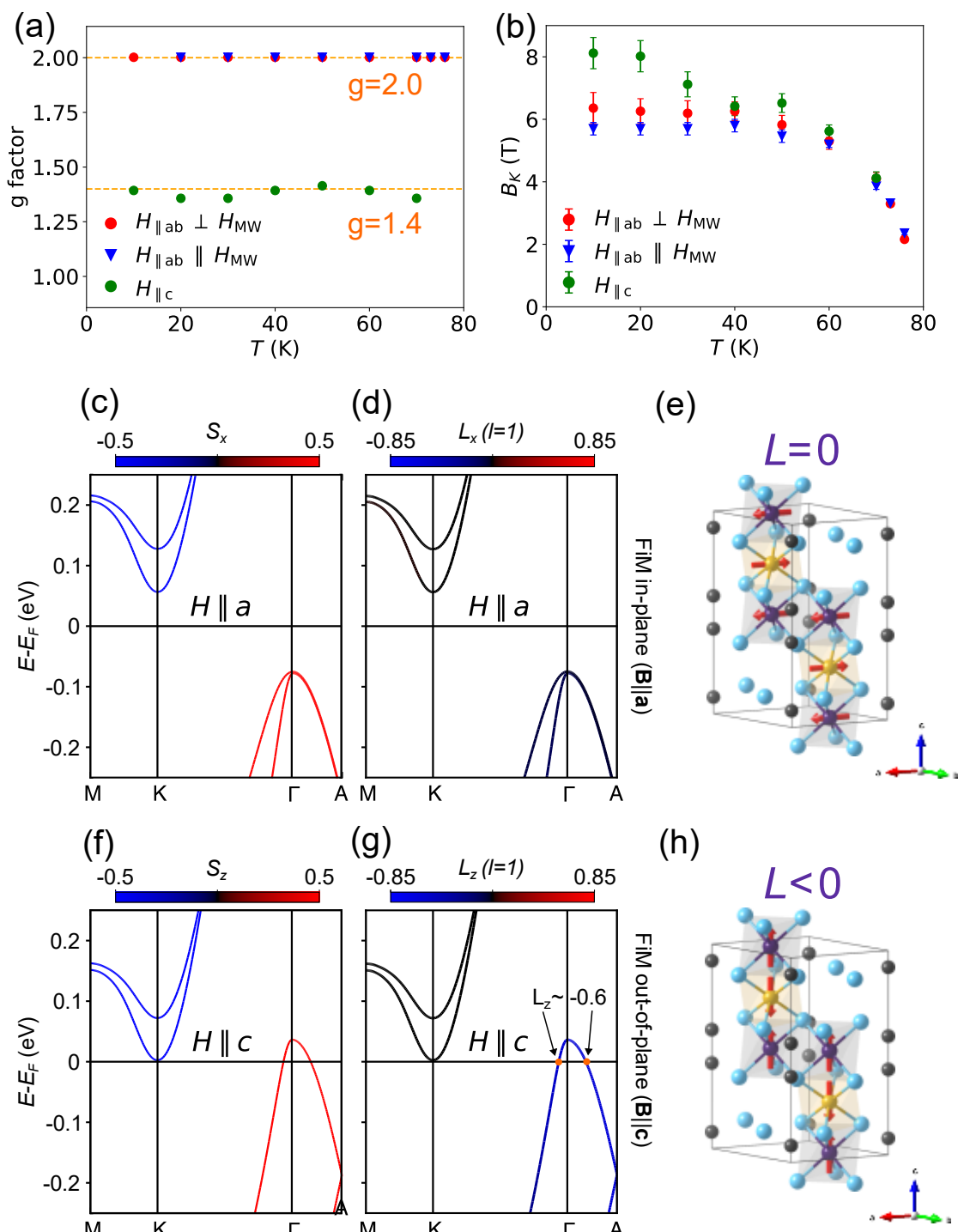

FIG 3. Anisotropy of g-factor and orbital magnetic moment. (a) g-factor as a function of temperature, obtained from the fitting of the resonance mode. The blue, orange, and green dots represent $H$||ab-plane, $H \perp H_{MW}$, $H$||ab-plane, $H$||$H_{MW}$, and $H$||c-axis configurations, respectively. (b) Anisotropy field Bk as a function of temperature, obtained from the fitting of the resonance mode. The blue, orange, and green dots represent $H$||ab-plane, $H \perp H_{MW}$, $H$||ab-plane, $H$||$H_{MW}$, and $H$||c-axis configurations, respectively. (c, d) Spin magnetic moment $S_x$ and orbital magnetic moment Lx, respectively, calculated for the electronic bands near the Fermi level, under the condition that the spins are aligned along the ab-plane. (e) Schematic of the spins aligned in ab-plane. (f, g) Spin magnetic moment $S_z$ and orbital magnetic moment $L_z$, respectively, calculated for the electronic bands near the Fermi level, under the condition that the spins are aligned along the c-axis. (h) Schematic of the spins aligned along c-axis and emergence of the orbital magnetic moment.

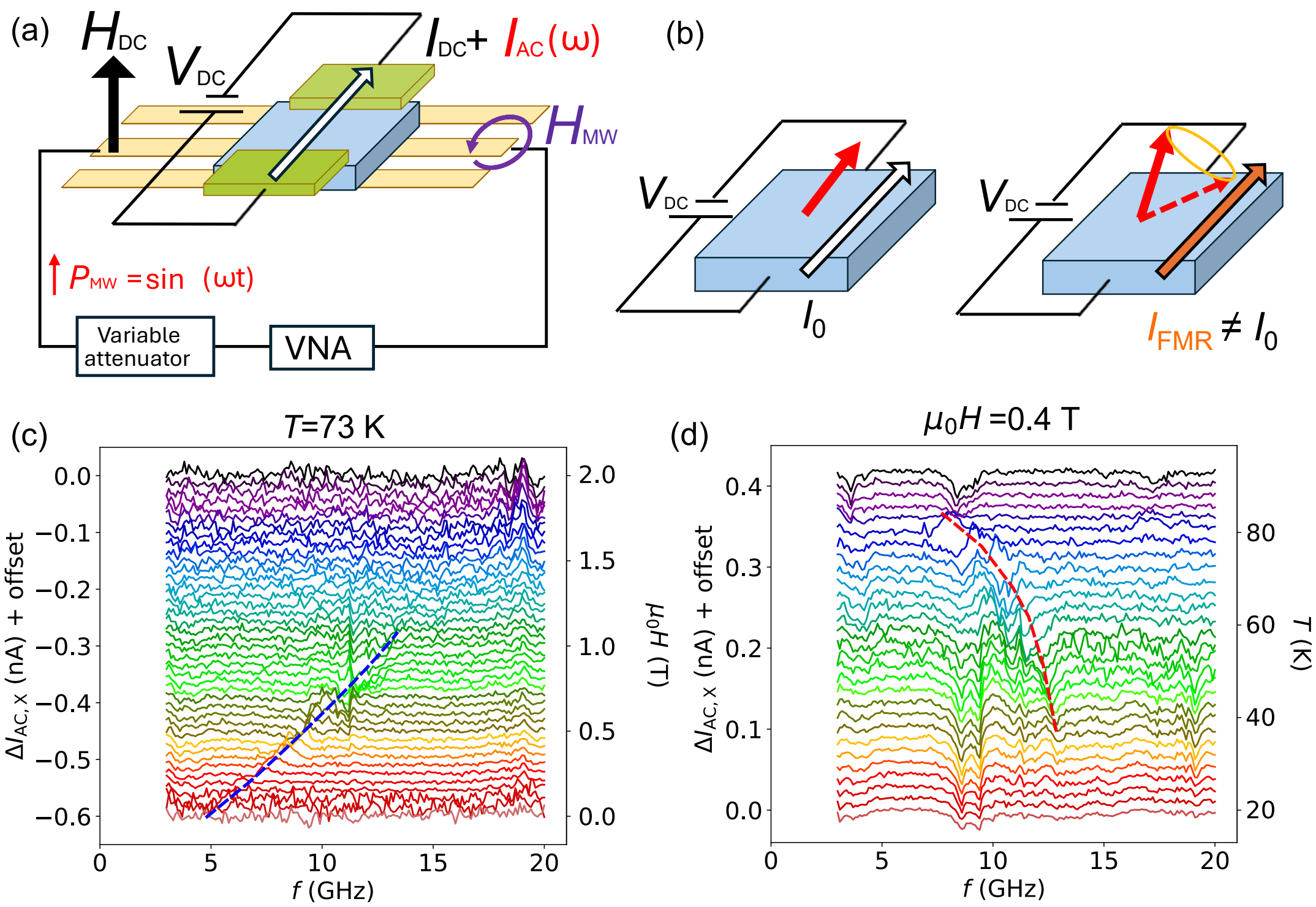

FIG 4. Resistance modulation under magnetic resonance. (a) Schematic of the lock-in measurement of microwave-induced resistance modulation. (b) Schematic of the canting of the spin and reduction of spin magnetization along the c-axis under the resonance. (c) Lock-in current modulation as a function of microwave frequency at magnetic field ($H$||c) from $H = 0$ T to 2 T (in 0.05 T interval) at $T$=73 K. The curves in different colors correspond to different magnetic fields and are vertically shifted for clarity. The magnetic field is indicated on the right axis. The blue dashed line serves as a guide to the eye for the resonance mode. (d) Lock-in current modulation as a function of microwave frequency at temperature from $T = 20$ K to 90 K (in 2.5 K interval) at $\mu_0 H = 0.4$ $T$ along c-axis. The curves in different colors correspond to different temperatures and are vertically shifted for clarity. The temperature is indicated on the right axis. The red dashed line serves as a guide to the eye for the resonance mode.